\documentclass[manuscript]{aastex}

\usepackage{subfigure}
\usepackage{longtable}

\newcommand{\II}{\scriptsize{II}\normalsize}
\newcommand{\I}{\scriptsize{I}\normalsize}

\slugcomment{}

\shorttitle{Chromospheric activity in AD Leo}
\shortauthors{Buccino et al.}

\begin{document}


\title{Possible chromospheric activity cycles in AD Leo }


\author{Andrea P. Buccino\altaffilmark{1,2,4}, Romina Petrucci\altaffilmark{1,4} , Emiliano Jofr\'e\altaffilmark{3,4},   Pablo J. D. Mauas\altaffilmark{1,2, 4}}

\altaffiltext{1}{Instituto de Astronom\'\i a y F\'\i
sica del Espacio (CONICET-UBA), C.C. 67 Sucursal 28, C1428EHA-Buenos
Aires, Argentina} \altaffiltext{2}{Departamento de F\'\i
    sica. FCEN-Universidad de Buenos Aires}
  \altaffiltext{3}{Observatorio Astron\'omico de C\'ordoba,
    C\'ordoba-Argentina} \altaffiltext{4}{Visiting Astronomer,
    Complejo Astron\'omico El Leoncito operated under agreement
    between the Consejo Nacional de Investigaciones Cient\'\i ficas y
    T\'ecnicas de la Rep\'ublica Argentina and the National
    Universities of La Plata, C\'ordoba and San Juan.}

\begin{abstract}

AD Leo (GJ 388) is an active dM3 flare star extensively observed both
in the quiescent and flaring states.

Since this active star is near the fully-convective boundary, to study
in detail its long-term chromospheric activity could be an appreciable
contribution for the dynamo theory.  Here, we analyze with the
Lomb-Scargle periodogram the Ca \II\- K line-core fluxes derived from
CASLEO spectra obtained between 2001 and 2013 and the V magnitude from
the ASAS database between 2004 and 2010. From both totally independent
time-series, we obtain a possible activity cycle of period $\sim7$
years and a less-significant shorter one of $\sim2$ years. A
tentative interpretation is that a dynamo operating near the surface
could be generating the longer cycle, while a second
  dynamo operating in the deep convection zone could be responsible
  for the shorter one.

Based on the long duration of our observing program at
CASLEO and the fact that we observe simultaneously different spectral
features, we also analyze the relation between simultaneous
measurements of the Na \I\- index ($R^{\prime}_D$), H$\alpha$ and Ca
\II\- K fluxes at different activity levels of AD Leo, including
flares.

\end{abstract}

{\it Object:}\object{GJ 388}
{\it Facilities:} \facility{ASAS}.

\section{Introduction}

Activity cycles like the one observed in the Sun have been detected in
several late-type stars
(e.g. \citealt{1995ApJ...438..269B,2008A&A...483..903B}). The usually
accepted model to describe the generation and intensification of
magnetic fields in these stars is the $\alpha\Omega$-dynamo first
invoked to explain solar activity \citep{1955ApJ...122..293P} and,
thoroughly improved and extended from F- to early-M stars (e.g.
\citealt{1982AA...108..322,1999ApJ...524..295S,2005ApJ...632.1104L}). These
dynamo models are based on the interaction between differential
rotation ($\Omega$-effect) and convective turbulence ($\alpha$-effect)
in the tachocline.

Cool stars with masses lower than about $0.35M_\odot$ are believed to
be fully-convective \citep{1997A&A...327.1039C}. Therefore, they do
not possess a tachocline and could not support an $\alpha\Omega$
dynamo. Nevertheless, there is plenty of observational evidence that
slow late-type rotators like dMe stars are very active and have strong
magnetic fields
(\citealt{1989PhDT.........1H,2004AJ....128..426W,2007ApJ...656.1121R}).
\cite{2006A&A...446.1027C} proposed that a pure $\alpha^2$-dynamo
could generate large-scale magnetic fields in fully-convective
stars. Moreover, the 3-D dynamo model  developed by
\cite{2008ApJ...676.1262B} for M dwarfs  reveals that fully-convective stars can
generate kG-strength magnetic fields without the aid of a shearing
tachocline.

To determine whether there is an onset of cyclic activity near the
convective limit, it is of special interest to study the long-term
chromospheric activity in stars of different spectral types, and in
particular in middle-M stars. As a contribution to this subject, since
1999 we have developed the HK$\alpha$ Project, an observing program
dedicated to periodically obtain mid-resolution echelle spectra of
southern late-type stars, including fully-convective ones. From our
data, we found evidences of cyclic activity for the fully-convective
M5.5Ve star Proxima Centauri \citep{2007A&A...461.1107C}, for the
mid-M spectroscopic binary GJ 375 \citep{2007A&A...474..345D} and the
two early-M stars GJ 229 A and GJ 752 A
\citep{2011AJ....141...34B}. Similarly, \cite{2012A&A...541A...9G}
studied a sample of stars from the \emph{HARPS} program and found that
the long-term activity of 8 early-M stars can be fitted by a
sinusoidal signal.

One of the most observed stars of our sample is AD Leo (GJ 388), which
is a M3Ve star \citep{1994AJ....108.1437H}, well-know for its frequent
\citep{1984ApJS...54..375P,2006A&A...452..987C} and strong flares
(e. g. \citealt{1991ApJ...378..725H}). Flares in AD Leo have been
observed and studied in the optical, EUV and X-ray
(e.g. \citealt{1995ApJ...453..464H,1996A&A...310..245M,2000A&A...354.1021F,2003ApJ...597..535H,2003ApJ...582..423G}).
Recently, \cite{2012PASP..124..545H} obtained a rotational period
$P_{rot}=2.23$ days for this star. AD leo is a partially-convective
star of mass $\sim0.40M_\odot$ \citep{2000A&A...354.1021F}.
\cite{2008MNRAS.390..567M} found that AD Leo hosts a magnetic field
with similar properties to several fully-convective stars. Although it
has been extensively observed both in the quiescent and flaring
states, most analysis of the chromospheric activity in AD Leo are
related to its short-term variability. Only \cite{1986A&AS...66..235P}
reported long-term variations in its flare frequency and suggest a
cycle in flaring activity.

This active fast-rotator star is an interesting target to explore the
dynamo around the threshold for full convection. Here we present a detailed
study of its long-term chromospheric activity. We delineate in
\S\ref{sec.obs} the main characteristics of the HK$\alpha$ Project and the
 ASAS program. In section \S\ref{sec.res}
we report our results, we analyze of the CASLEO activity
indicators and the long-term magnetic activity derived from
CASLEO data and the ASAS database. Finally, we discuss our results in the
context of the dynamo theory.

\section{Observations}\label{sec.obs}

To study the long-term chromospheric activity in late-type stars,
in 1999 we started the HK$\alpha$ Project. In this program, we
systematically obtain mid-resolution echelle spectra ($R=\lambda/\delta\lambda\approx26000$) of several late-type stars.  Our
observations are made at the 2.15~m telescope of the \emph{Complejo
  Astron\'omico El Leoncito (CASLEO)}, in San Juan, Argentina. Specific
details of the observations and the method used to calibrate the
spectra is described in \cite{2004A&A...414..699C}.

 Our stellar sample was chosen to cover the spectral range from F to
 M, with different activity levels. In particular, we included a large
 number of M stars to study the transition to the completely convective
 regime. Most of the stars are single dwarfs, although we also include several
 binaries and a few subgiants.  At present, we have about 5500 spectra
 of 150 stars.

The standard activity indicator used in dF to dK stars is the
Mount Wilson $S$ index, essentially the ratio of the Ca \II\- H and K
line-core fluxes to the continuum nearby. However, as explained in detail in
\cite{2011AJ....141...34B}, it is not suitable to study the
chromospheric activity on AD Leo. Following our previous works
\citep{2007A&A...474..345D,2011AJ....141...34B}, we use as a proxy of
stellar activity the Ca \II\- K line-core flux, integrated with a
triangular profile of 1.09~\AA\- FWHM centered in 3933.66~\AA\,
\citep{2007astro.ph..3511C}.

We complement our data with photometry from the All Sky Automated
Survey\footnote{\textsf{http://www.astrouw.edu.pl/asas/}} (ASAS). The
ASAS program systematicallyobserves around 10$^7$ stars brighter than
V=14, simultaneously in the V and I bands. Here we use the V
magnitude to analyze the long-term activity of AD Leo.

\section{Results}\label{sec.res}
\subsection{Chromospheric activity cycles}\label{sec.gl388casa}
In Table \ref{tab.logs} we show the observation logs of AD Leo at
CASLEO.  There is a total of 38 individual observations, which have
been carried out on 19 nights distributed over 12 years between March
2001 and March 2013. Our observations consist of two successive
spectra, to eliminate cosmic rays. Observation times are between 30
and 60 minutes.

\cite{2006A&A...452..987C} observed
a large number of short and weak flares occurring very frequently
(flare activity $>0.71$~h$^{-1}$), which can mask the variations of
activity due to the cycle. Therefore, before exploring the existence
of an activity cycle in this star, we filter out any flares
from our observations. We do it by visual inspection of the
observations, since during flares the line fluxes in the two succesive
spectra are very different. We
excluded these flaring spectra, which are indicated in Table
\ref{tab.logs} with a ``$\star$'', from the rest of the analysis.

\begin{figure}
\subfigure[\label{fig.gl388cas}]{\includegraphics[width=0.5\columnwidth]{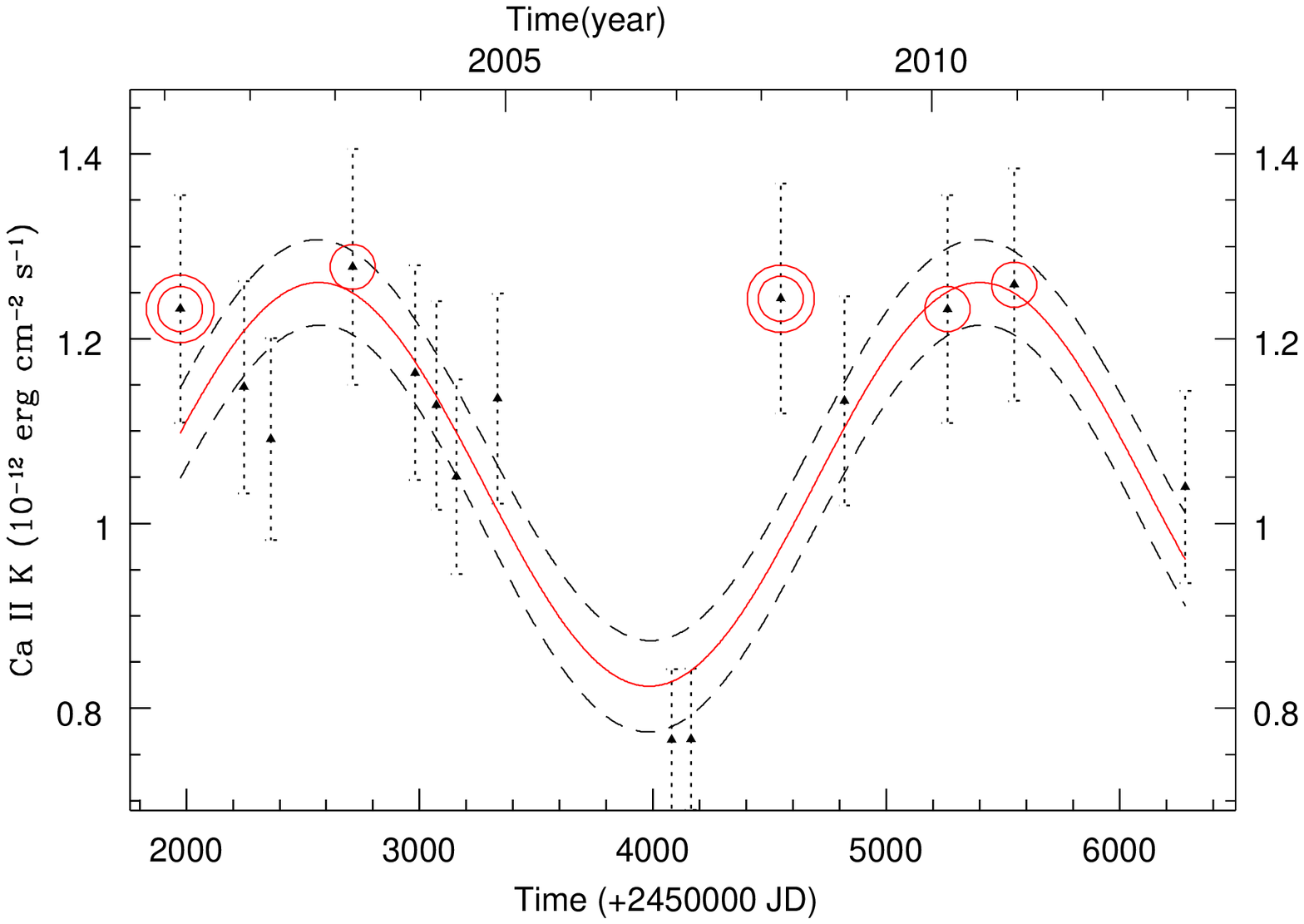}}\hfill
\subfigure[\label{fig.gl388as}]{\includegraphics[width=0.5\columnwidth]{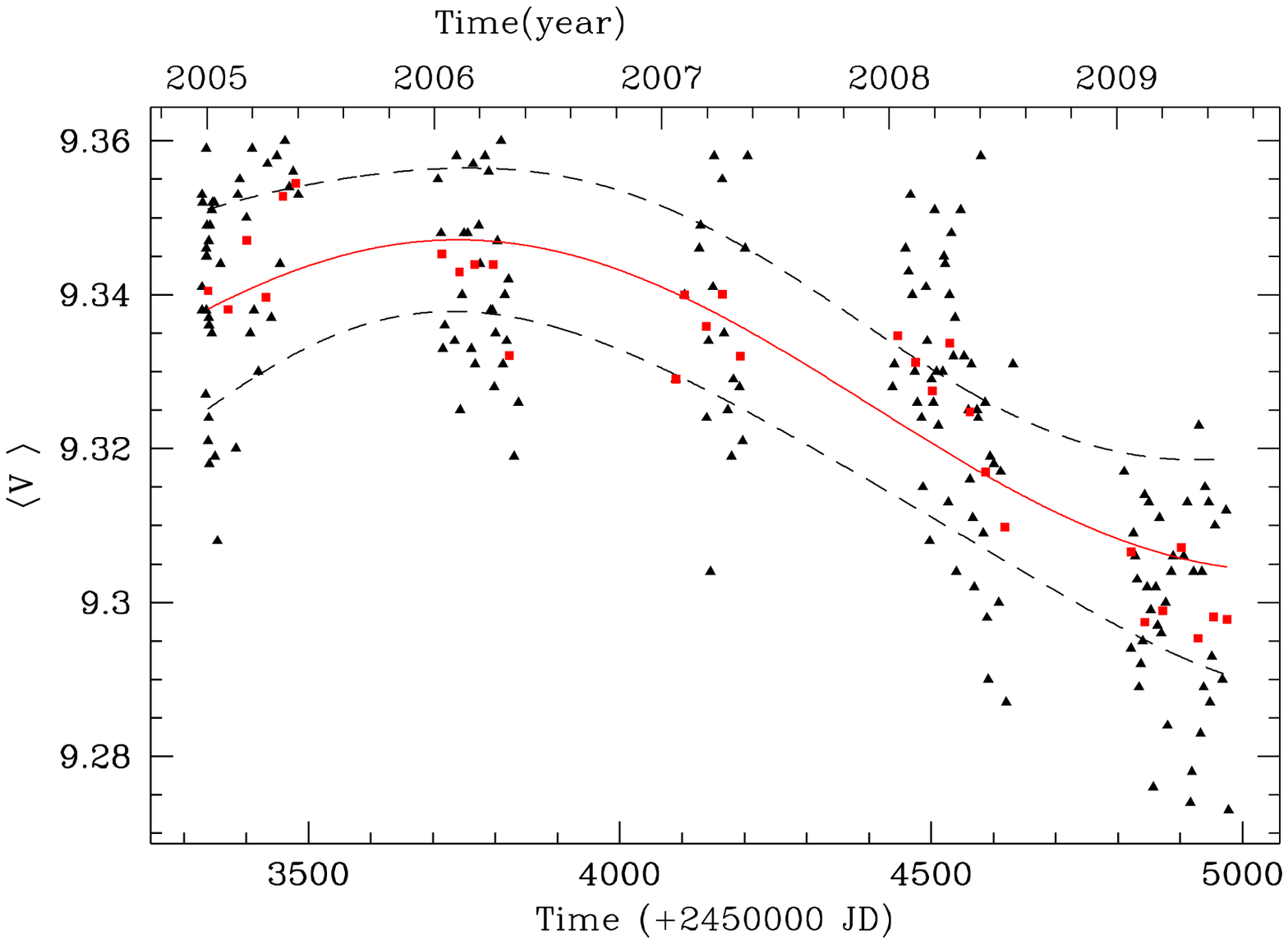}}
\caption{(a) Ca \II\- K fluxes derived from the non-flaring CASLEO
  spectra, assuming errors of 10\% \citep{2004A&A...414..699C}. The
  solid line is the least-square fit with a harmonic function of
  period 2769 days, the most signifcant peak in
  Fig. \ref{fig.gl388per}, with $\chi_r=0.78$. The dashed lines
  represent $\pm$1$\sigma$ deviations. The red circles indicate those
  points associated with high activity, more than 0.5-$\sigma$ the
  mean. Removing the outlier points (double circled),
    which might be affected by weak flares, we obtain
    $P_{1,CASLEO}=2793$ days with a better FAP=0.3\%. (b) V magnitude
  measured by ASAS. The solid line is the least-square fit with a
  harmonic curve of period 2659 days ($\chi_r=0.49$), as obtained in
  Fig. \ref{fig.gl388per}. The dashed lines represent those points
  which appart in 1-$\sigma$ from this fit. The monthly means are
  indicated with square points.}\label{fig1}
\end{figure}

For the non-flaring CASLEO spectra, we calculated a nightly average of
the Ca \II\- K fluxes. We plot the resulting time-series of these
activity indicators in Fig.~\ref{fig.gl388cas}. In this figure we
observe that AD Leo reaches the minimum of activity (40\% lower than
at the maximum) at the year 2007.

On the other hand, we obtained the V magnitude of AD Leo from the ASAS catalog
for the period between 2004 and 2010.  We only included the best
quality data (see \citealt{2011AJ....141...34B}) and  we discarded 14 outlier observations. In
Fig. \ref{fig.gl388as} we plot the resulting time-series, which
consists of 175 points. Typical errors are around 30 mmag. The mean
magnitude of the whole dataset is around $\langle V\rangle=9.327\pm0.021$.  We also look for evidence of rotational modulation due to
spots and active regions on the stellar surface, probably responsible
for the short-scale variations ($\sim$0.5\%). To do so, we analyzed
the data of each observing season with the Lomb-Scargle periodogram
\citep{1986ApJ...302..757H}. Nevertheless, we did not detect any significant
periodicity in the ASAS seasonal datasets; probably due to the
low-precision of the ASAS photometry.

\begin{figure}
\subfigure[\label{fig.gl388per}]{\includegraphics[width=0.5\columnwidth]{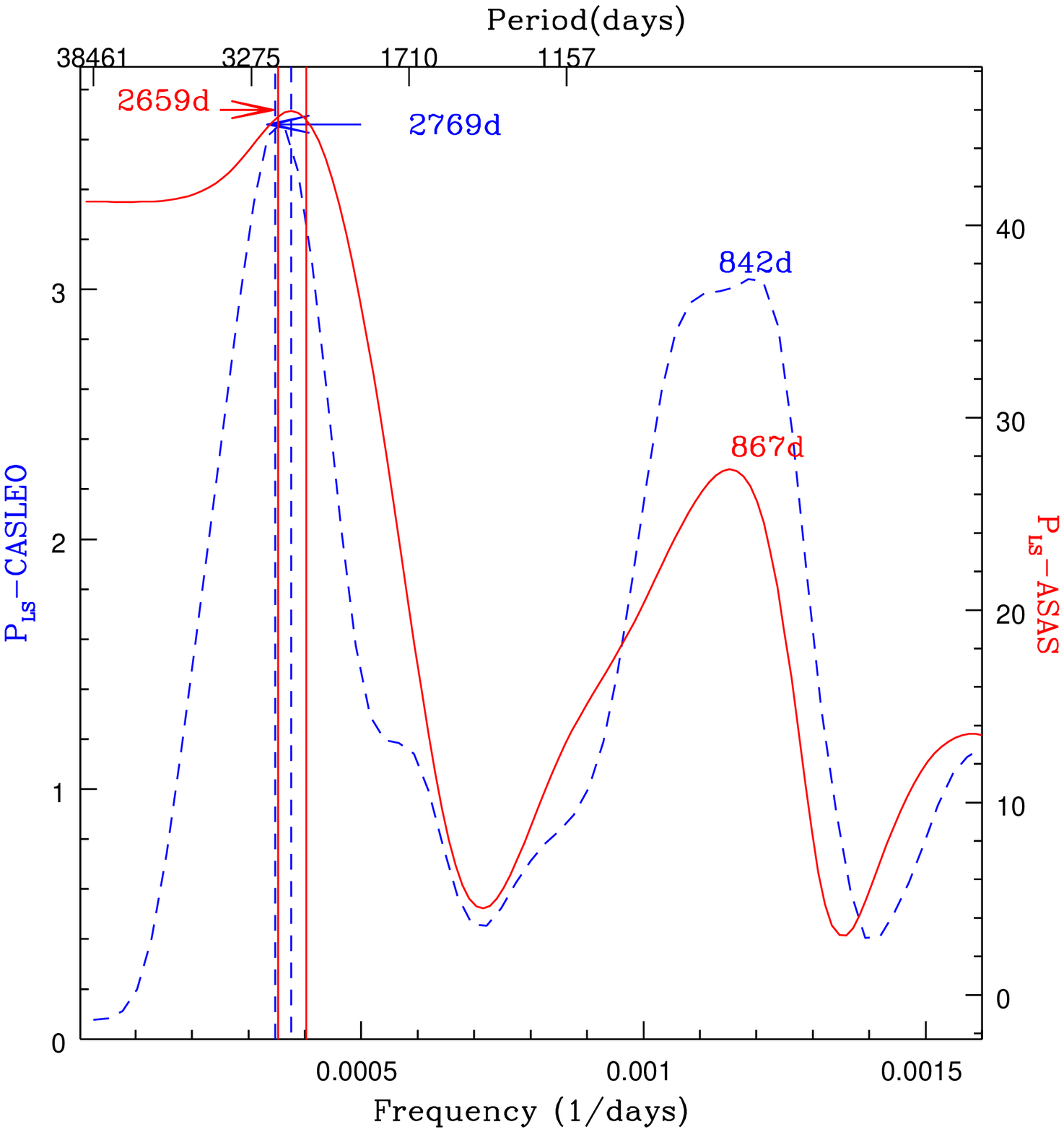}}
\subfigure[\label{fig.gl388casas}]{\includegraphics[width=0.5\columnwidth]{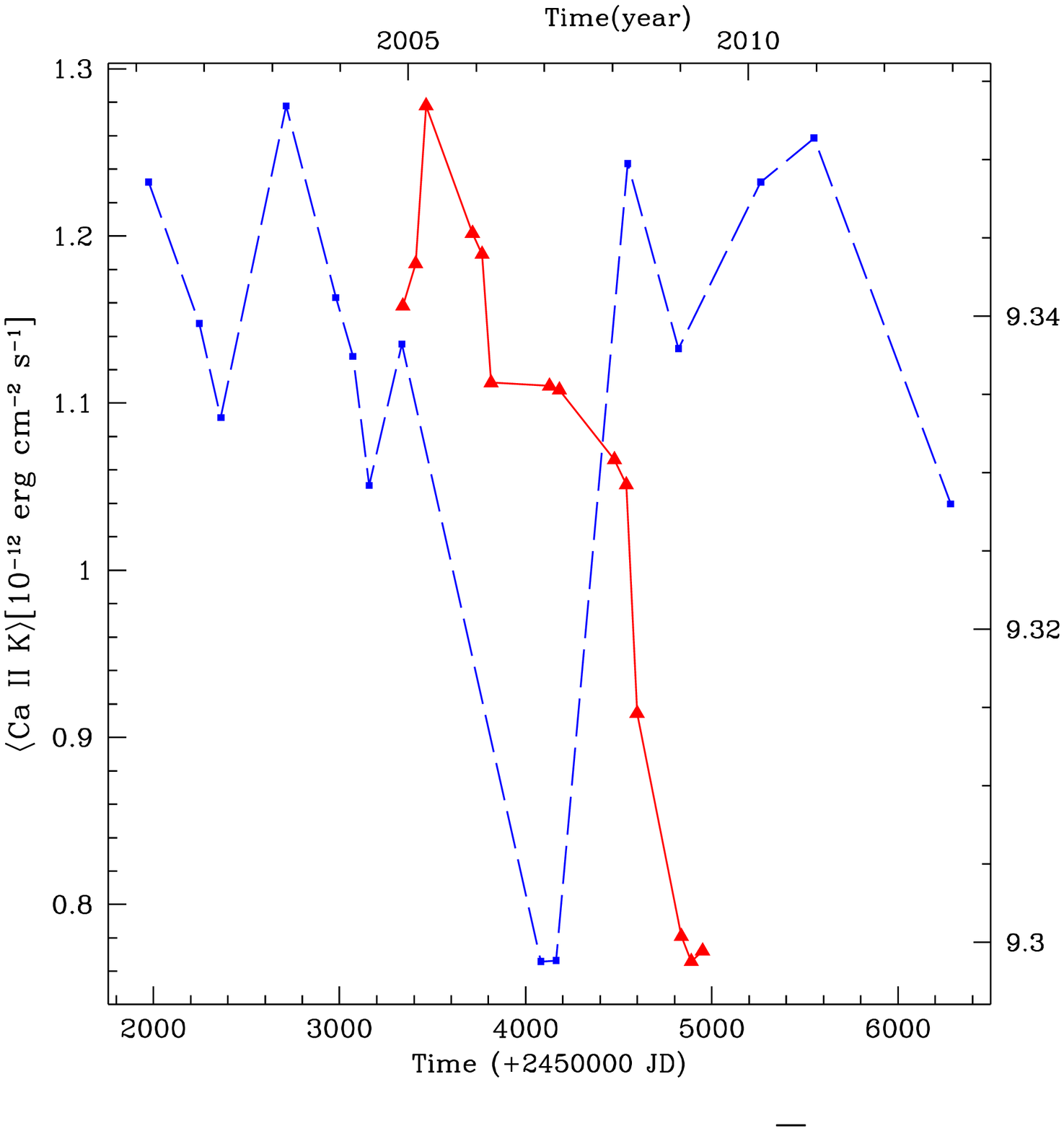}}
\caption{ (a) Lomb-Scargle periodogram of the Ca \II\- K fluxes
  (dashed) and the V magnitude (solid). We represent with vertical
  solid and dashed lines the error interval of each period. (b) The Ca
  \II\- K fluxes plotted in Fig.~\ref{fig.gl388cas} (dashed) and the quarterly weighted mean  V magnitude obtained from the data plotted in
  Fig.~\ref{fig.gl388as} (solid). Typical errors are of $\sim10\%$ for
  the Ca \II\- K fluxes and $<0.12\%$ for the mean V magnitudes. For
  clarity, we do not include the error bars.  }\label{fig2}
\end{figure}

 To search for long-term chromospheric activity cycles, we first
analyzed both independent datasets with the Lomb-Scargle
periodogram. The False Alarm Probability (FAP) of the periods obtained
were computed with a Monte Carlo simulation, described in \cite{2009A&A...495..287B}.  The periodogram obtained for the CASLEO
data is plotted in Fig. \ref{fig.gl388per} with a dashed line. It
shows a primary periodicity in the Ca \II\- K fluxes with a period
$P_{1,CASLEO}=(2769\pm106)$ days with FAP=8\%, and a second, less
significant, harmonic component with period $P_{2,CASLEO}=(842\pm9)$
days with FAP=30\%.  To analyze the robustness of our results, we
computed the periodograms of the series obtained by eliminating each
data point alternatively. In 11 of 14 periodograms we obtained 
$P_{1,CASLEO}$ between 2593 and 2982 days with FAPs$<$30\% (73\% of
them were $P_{1,CASLEO}=2769$ days) and a secondary peak
$P_{2,CASLEO}$ between 843 and 931 days with larger FAPs.

To investigate whether the 842-day period is a subharmonic frequency
of the prominent peak, we used the monthly sunspot numbers ($S_N$)
taken from the National Geophysical Data Center\footnote{
  \textsf{http://www.ngdc.noaa.gov/stp/solar/ssndata.html}} to
  take into account that activity does not follow a strictly sinusoidal
  law. We used $S_N$ between 1751 and 2012 and rescaled the series in
time to the period $P=2769$ days ($\sim$7.58 years). We also rescaled
the $S_N$ to obtain a time series of the same mean value and standard
deviation as our data.  To consider data with the same signal-to-noise
as ours, we added Gaussian noise with errors of 10\% at each point. We
took 1000 samples of these data with random starting dates and the
same phase intervals that we have in our data and we computed the
periodograms. As expected, 55\% of the most significant periods
detected in each periodogram appart from $P=2769$ days in less than
10\%. On the other hand, only 0.3\% are between 671 and 1007 days (842
days $\pm20\%$).  Therefore, we can conclude that the peak at $P_{2,
  CASLEO}=842$ days ($\sim$2.3 years) is not an
artifact. We also performed a similar analysis on a
  sinusoidal function of period 2769 days with 10\%-Gaussian
  noise. While 82\% of the periods detected are 2769 days ($\pm$10\%),
  periods near $P_{2, CASLEO}$ were not detected. This analysis
  confirms that this secondary peak is not due to aliasing.

In Fig. \ref{fig.gl388per} we show the periodogram for the ASAS time
series. We obtained prominent peaks at $P_{1,ASAS}=(2569\pm107)$ days
(FAP$\sim 10^{-12}$) and $P_{2,ASAS}=(867\pm6)$ days (FAP$\sim
10^{-9}$). \textbf{Since the ASAS data timespan is only $\sim$4.5
  years,  periods longer than $P_1$ are also potentially significant,
  as indicated by high periodogram power (Fig.~\ref{fig.gl388per}).} The large number of points of this series
are responsible for these extremely low FAPs (see
\citealt{1986ApJ...302..757H}, Eq. 22). To check that these periods
are significantly independent of the data dimension, we reduce the
number of points by binning the data. We computed the monthly mean V
magnitude \textbf{(i.e. averaging on timescales much longer than
  $P_{rot}$)}, weighted by the error reported in the ASAS database,
and we computed the error of each mean magnitude as the square root of
the variance-weighted mean (see \citealt{Frod79}, Eq. 9.12). For this
series we detected a period $P_{ASAS}=(2937\pm575)$ days, with a
FAP=$10^{-5}$. Since we smoothed the data plotted in
Fig. \ref{fig.gl388as}, the secondary period near $\sim$900 days is
much less significant (FAP=5\%) in this periodogram.

The Ca \II\- K fluxes and the mean quarterly magnitudes are plotted
together in Fig.~\ref{fig.gl388casas}. Contrary to the Sun, we observe
that spots dominate the emission in AD Leo as it becomes fainter when
the Ca \II\- emission increases. On the other hand, there is an
evident timelag between both chromospheric series. Both datasets
coincide within the normalization constant with a correlation
coefficient R=0.95, if we shift the photometry by 770 days.

\cite{1995ApJ...441..436G} and
\cite{1996ApJ...465..945G,1996ApJ...456..365G} have already observed
this timelag between photometric and magnetic variations for stars of
different spectral types (G0V-K2V).  \cite{1996ApJ...465..945G} showed
that, when different stars are compared, this timelag is
anti-correlated with effective temperature. However, the Sun does not
fit this relation \citep{1997ApJ...474..802G}. Furthermore, here we
find that AD Leo does not follow this trend, similarly to what we found for other M-stars
\citep{2007A&A...474..345D,2011AJ....141...34B}. The physical
explanation for these timelags remains unknown
\citep{2008ApJ...679.1531B}.

\subsection{H$\alpha$ and Na I D lines as activity indicators}\label{subsec.act}

Due both to their red color and faint intrinsical luminosity, it is
quite difficult to observe the Ca \II\- lines in dM stars with
adequate signal-to-noise , specially for programs aiming to follow a
large number of stars. Therefore, it would be very convenient to find
other activity indicators at longer wavelengths. Since we observe
simultaneously a long range of wavelenghts, our data provides an
excellent opportunity to study the correlation between different
spectral features and activity indexes. Furthermore, the long duration
of our observing program, the HK$\alpha$ Project, ($\sim$14 years)
allows us to analyze if the relation between these indexes depends on
the level of activity of each individual star and, therefore, if it is
associated to the distribution of active regions in the stellar
atmosphere.

\cite{2007MNRAS.378.1007D} studied the Na \I\- D lines
(D$_1:5895.92$ \AA; D$_2:5889.95$ \AA) in our stellar sample with CASLEO
spectra. They constructed a spectral index ($R^\prime_D$) as the ratio
between the flux in the D lines and the bolometric flux. They
concluded that this index, once corrected for the photospheric
contribution, can be used as a chromospheric activity indicator in
stars with H$\alpha$ in emission. Using high-resolution HARPS
spectra, \citet{GdS11} were able to integrate the D line flux in
narrower windows, and, in 70\% of their sample of M stars, they found
a significant correlation with the Ca flux, even at low activity
levels. In particular, they always obtained  a positive correlation for
each individual star.

\begin{figure}
\subfigure[\label{fig.gl388naca}]{\includegraphics[width=0.5\columnwidth]{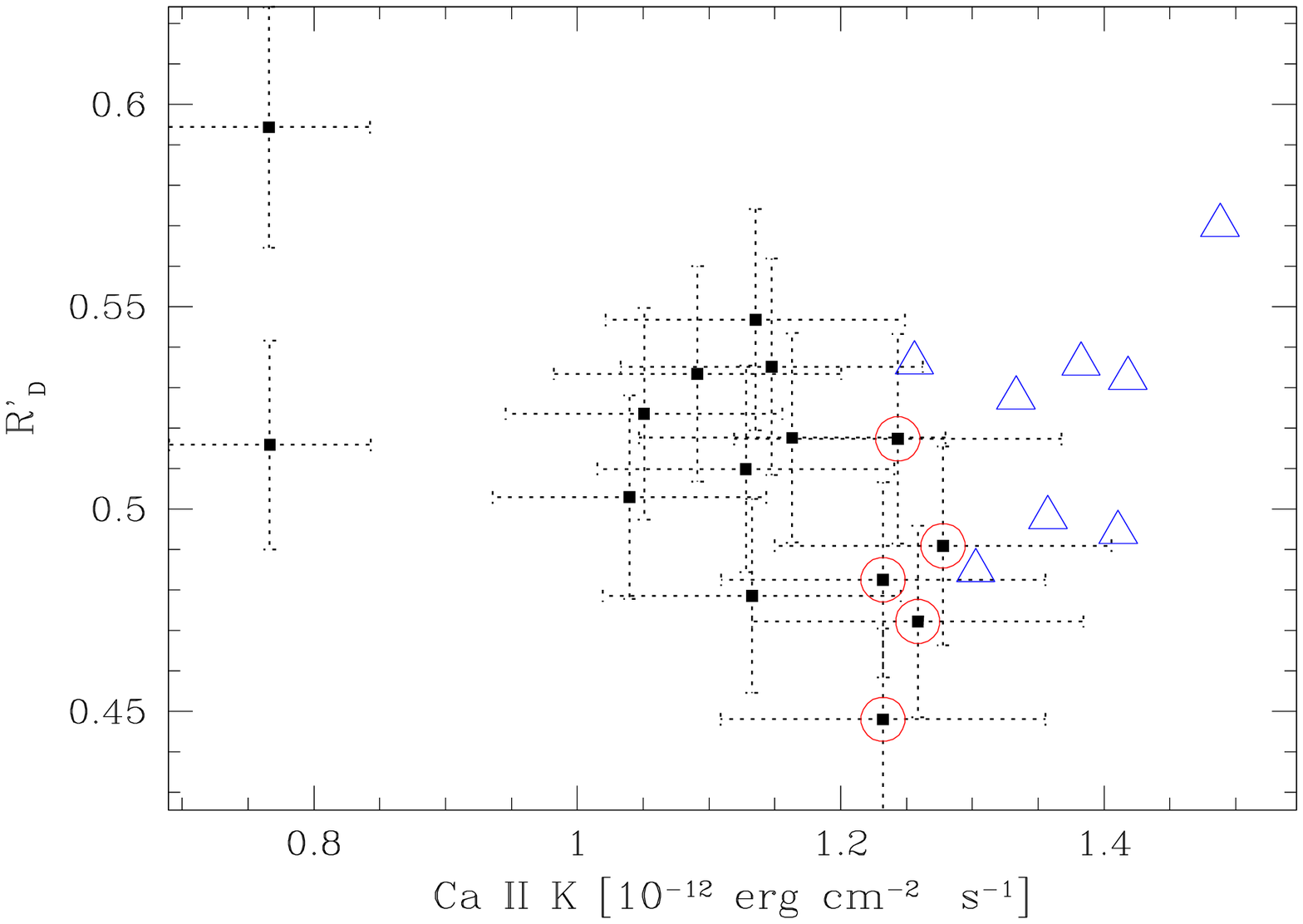}}\hfill
\subfigure[\label{fig.gl388haca}]{\includegraphics[width=0.5\columnwidth]{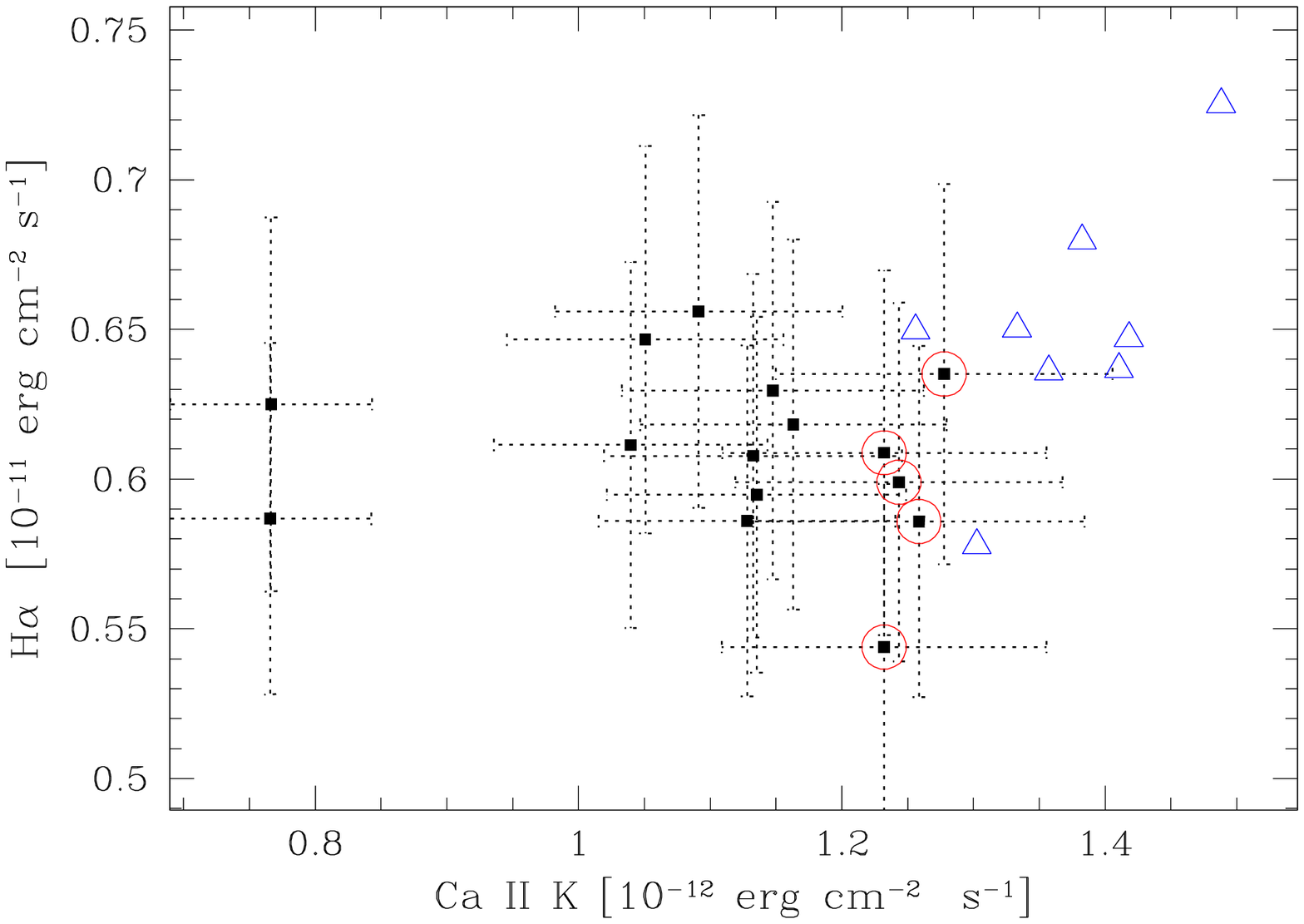}}
\caption{Simultaneous measurements of the Na \I\- index $R^{\prime}_D$ (a) and the H$\alpha$ line-core flux (b) vs. the Ca \II\- K fluxes from CASLEO spectra. The red circles indicate those points  where the level of  activity  exceeds in more than 0.5-$\sigma$ the mean, and the triangles signal those points associated with flares.}\label{fig3}
\end{figure}

In Fig. \ref{fig.gl388naca} we plot simultaneous measurements of the
$R^\prime_D$ index vs. the Ca \II\- K fluxes derived from CASLEO
spectra of AD Leo. For the non-flaring points, both indexes
anti-correlate with a Pearson's correlation coefficient
R=-0.63. If we remove the maximum $R^\prime_D$, a possible
  anticorrelation is still present, but much less significant
  (R=-0.39). This tendency changes during flares when both activity
indicators correlate with R=0.43.

Although in the Sun the correlation between the Ca \II\- K and H$\alpha$
line-core fluxes is positive \citep{2007ApJ...657.1137L}, \cite{2007astro.ph..3511C} reported that
this relation is not always valid for other late stars (F7V-M5V).
Each star shows a particular behaviour, ranging from anti-correlations
to tight correlations with different slopes, including cases where no
correlations are found \citep{2007astro.ph..3511C}. Based on this
result, \cite{2009A&A...501.1103M} studied the H$\alpha$-Ca \II\-
relation during the solar cycle and they found that this correlation
and the slope were much larger during solar maximum than during
minimum.  In \cite{2012IAUS..286..324B}, we show simultaneous
measurements of the H$\alpha$ and Ca \II\- K+H fluxes for three solar
analogs of different ages. Although they show a low
correlation for the whole series, the correlation is strongly positive
during active phases.

In several M stars of our sample we found that
H$\alpha$ and the Ca \II\- K fluxes were not correlated
\citep{2007A&A...474..345D,2011AJ....141...34B}.\cite{2009AJ....137.3297W} studied this correlation for
    several M3 dwarfs, using one spectrum for each star, and found
    a strong positive correlation between simultaneous measurements
    of Ca \II\- K and H$\alpha$ for the most active stars, with
    H$\alpha$ in emission, including AD Leo. However,
    they did not analyzed this relation for individual observations of
    each star.
  
 Here, we compute the H$\alpha$ flux as the average surface flux in a
 1.5 \AA\- square passband centered in 6562.82 \AA
 \citep{2007astro.ph..3511C}. In Fig. \ref{fig.gl388haca} we plot the
 H$\alpha$ and Ca \II\- K fluxes for the spectra used in
 Fig. \ref{fig.gl388cas}. Both fluxes are not correlated
 (R=-0.145). However, as we found for solar-analogs, those points
 associated with high activity show a rather strong correlation, with
 coefficient R=0.64, and are also
 correlated during flares, with R=0.65. These points are shown as
 triangles in Fig. \ref{fig.gl388haca}.

\section{Discussion}
To explore the dynamo near the fully-convective boundary, we studied
the long-term activity of the M3Ve dwarf star AD Leo (GJ 388). In
particular, we analyzed the Ca \II\- K line-core fluxes measured on
our CASLEO spectra obtained since 2001, and the ASAS photometric
data. Both in the photometry and in the Ca \II\- fluxes, we obtained a
possible activity cycle of $\sim$7 years and a less-significant
shorter one of $\sim$2 years with the Lomb-Scargle periodogram.  It
should be noted that we detected similar periods in two completely
independent datasets. These reinforces the significance of the
detection, since the probability of a false detection in both datasets
should be computed as the product of both FAPs. The longer cycle
coincides whithin the statistical errors whith the periodic variations
in the flare frequency (for energies above 10$^{30}$ ergs) reported by
\cite{1986A&AS...66..235P}. Furthermore, the minimum of the $\sim$7-yr
cycle reached in 2007 also coincides with a significant decrease in
flare activity \citep{2008RoAJ...18S..55K}.

Multiple cycles have been already detected in several cool stars
(e.g. \citealt{1999ApJ...524..295S,2009A&A...501..703O,2013ApJ...763L..26M}).
For those stars later than F5 with well-determined rotation and cycle
periods, \cite{2011IAUS..273...61S} examined the relation between the
cycle ($\omega_{cyc}$) and rotation frequencies ($\Omega$), and found
that most cycles fall into three parallel tracks, with
$\omega_{cyc}\sim\Omega^{1.1}$, with double cycles falling in
different branches. Therefore, both periods are not independent. In
fact, given the primary period of AD Leo of around 2800 days and its
rotation period of 2.23 days, Saar's results imply that if this star
presents a secondary cycle, its period should be around 800 days,
similar to the one we obtain.  A possible interpretation for this
bimodal relation is that each activity cycle has different physical
sources \citep{2007ApJ...657..486B}. Probably two dynamos may
  be operating inside the star: one driven by rotational shear in the
  near-surface layers (longer cycle) and the other in the deep
  convection zone (shorter cycle). Since AD Leo is a
  \textbf{near fully-convective} star, unlike the stars considered in
  \cite{2007ApJ...657..486B}, a tachocline dynamo may not be feasible
  in this star.

Since we have simultaneous observations of different spectral features
we can analyze the relation between chromospheric lines formed at
different depths along the activity cycle and during flares. In
particular, we found evidences of a possible anticorrelation
between the Na \I\- index and the Ca \II\- K fluxes. On the other
hand, the H$\alpha$ and Ca \II\- fluxes show little correlation during
the cycle. Since the observed fluxes add the radiation coming from
different magnetic structures, like spots and faculae, the resulting
fluxes change during the cycle as the filling factors for each of
these structures change. Therefore, the correlation between fluxes can
depend on each line's contrasts (see
  \citealt{1996A&A...310..245M}).

In particular, during flares we find a positive correlation between
the Ca \II\-, Na \I\- and H$\alpha$ fluxes, suggesting that the
flare-heating mechanism in operation from mid- to high-chromosphere
does not change with flare strength, suggested by
\cite{2003ApJ...597..535H}.

M dwarfs are ideal targets to search for terrestrial planets in the
habitable zone. However, their activity signatures can hinder the
detection of orbiting planets. Our results suggest that the level of
activity will be appreciably lower during the next minimum, expected
around 2015. This should be the best moment to search for planets
orbiting AD Leo.

\clearpage




\clearpage







\begin{deluxetable}{ccc}
\tabletypesize{\scriptsize}
\tablecaption{Logs of the CASLEO observations of AD Leo.\label{tab.logs}}
\tablewidth{0pt}
\tablehead{
\colhead{Label$^\textrm{a}$} & \colhead{xJD$^\textrm{b}$} & \colhead{t$^\textrm{c}$} 
}
\startdata
0301c1 & 1974.57 &    2700 \\
0301c2 & 1974.61 &    2700\\
$\star$1201c1 & 2247.82 & 1800\\
1201c2 & 2247.84 & 1800\\
0302a1 & 2363.61 & 1800\\
0302a2 & 2363.63 & 1800\\
0303a1 & 2713.63 & 3000\\
0303a2 & 2713.66 & 3000\\
1203b1 & 2980.82 & 1800\\
1203b2 & 2980.84 & 1800\\
0304a1 & 3072.62 & 3000\\
0304a2 & 3072.66 & 3000\\
0604b1 & 3159.47 & 1800\\
$\star$0604b2 & 3159.49 & 1800\\
1104d1 & 3335.82& 1800\\
1104d2 & 3335.84 & 1800\\
$\star$0305a1 & 3448.64 & 1800\\
$\star$0305a2 & 3448.67 & 1800\\
1206d1 & 4081.83 & 2400\\
1206d2 & 2980.86 & 2400\\
0307a1 &4162.69 & 3600\\
0307a2 &4162.73 & 3600\\
0308b1 & 4547.64    &   3600\\
0308b2 & 4547.68 & 3600\\
1208c1 & 4821.77& 3600\\
1208c2  & 4821.82& 3600\\
$\star$0309b1 &4902.63 &3600\\
0310b1 &5262.58& 3600\\
0310b2 &5262.62& 3600\\
1210c1 &5547.77 & 2700\\
1210c2 &5547.81 & 2700\\
1212b1 &6281.82 & 1800\\
1212b2 &6281.84 & 1800\\
$\star$0313b1&6354.64 & 2850\\
$\star$0313b2& 6354.67 & 2850\\
\enddata
\tablenotetext{a}{\,Label built with month and year of the observation (e.g. 0305 refers to March, 2005), a letter for different nights of the same run and a number for each individual nightly spectrum.}
\tablenotetext{b}{xJD=JD-2450000, where JD is the Julian date when the observation begins. }
\tablenotetext{c}{Exposure time in seconds.}
\tablecomments{Spectra indicated with $\star$
are discarded in Fig. \ref{fig.gl388cas}, due to the presence of flares.}
\end{deluxetable}

\bibliographystyle{apj}

\end{document}